\documentclass[12pt]{article}

\usepackage[totalheight = 23cm, totalwidth = 17cm]{geometry}
\usepackage{amssymb,amsmath,amsfonts,amsbsy,graphicx,bm}

\usepackage{color,cancel,ulem}

\newcommand{\dis}[1]{\begin{equation}\begin{split}#1\end{split}\end{equation}}

\begin{document}

\begin{titlepage}

\begin{center}

{\LARGE \bf 
 Axion species scale and axion weak gravity conjecture-like bound
}

\vskip 1.0cm

{\large
Min-Seok Seo$^{a}$ 
}

\vskip 0.5cm

{\it
$^{a}$Department of Physics Education, Korea National University of Education,
\\ 
Cheongju 28173, Republic of Korea
}

\vskip 1.2cm

\end{center}

\begin{abstract}

As a cutoff scale of quantum gravity, the species scale can be defined by the scale at which the perturbativity of the non-renormalizable gravitational interaction begins to break down.
Since it is determined by the number of  species in the effective field theory, we can find the close connection to the distance conjecture, which predicts the lowering of the cutoff at the asymptotic limit of the moduli space  caused by the descent of a tower of states from UV.
Meanwhile, the same kind of the cutoff scale can be obtained from any non-renormalizable interaction, in particular the interaction between the axion and the gauge field through the $\theta F\wedge F$ term.
Demanding this `axion species scale' not to exceed the gravitational species scale, we obtain the bound $(8\pi^2/g^2)f \lesssim M_{\rm Pl}$.
This is quite similar  to the axion weak gravity conjecture bound, but  can be applied to any gauge as well as the string  interactions which are relevant to  towers of states.
We also investigate the implications of the (axion) species scale and the axion weak gravity conjecture-like bound by considering the   Peccei-Quinn charge reduction of  black hole through the interaction between   black hole and the string or wormhole.

\end{abstract}

\end{titlepage}

\newpage

\section{Introduction}

 The Axion is the pseudoscalar field with a perturbative global symmetry called the Peccei-Quinn (PQ) symmetry  $\theta \to \theta+c$, where $\theta$ is the axion and $c$ is some constant   (see, e.g., \cite{Kim:1986ax, Kim:2008hd} for reviews).
 It can be regarded as the dynamical promotion of the CP violating theta angle since it couples to the gauge field strength $F$ through the interaction term $\theta F\wedge F$.
 From this, the axion is predicted to resolve the strong CP problem in a dynamical way \cite{Peccei:1977ur, Peccei:1977hh, Weinberg:1977ma, Wilczek:1977pj}.
 More concretely, the non-perturbative QCD instanton breaks the PQ symmetry and induces the axion potential.
 Then  $\theta$ is stabilized at zero, resulting in the CP conservation in QCD \cite{Vafa:1984xg}. 
 Whereas two types of the axion,  the KSVZ \cite{Kim:1979if, Shifman:1979if} and  DFSZ \cite{Zhitnitsky:1980tq, Dine:1981rt} models, were proposed in the context of quantum field theory without gravity, it was realized that there can be plenty of axion-like particles in string compactifications \cite{Witten:1984dg} (see  \cite{Svrcek:2006yi} and references therein).
 Some of such `stringy axions' may be identified with the QCD axion  (see, e.g., \cite{Burgess:1998px, Choi:2003wr, Conlon:2006tq, Choi:2014uaa, Gendler:2024gdo}), and even if irrelevant to the strong CP problem,  they also can be used to explain various phenomena which are yet to be understood. (see, e.g., \cite{Arkani-Hamed:2003xts, Svrcek:2006hf, Arvanitaki:2009fg}, and \cite{Choi:2020rgn} for a recent review).

 Meanwhile, in string theory, the reliable low energy effective field theory (EFT) can be constructed by taking the asymptotic limit of the moduli space.
 In this regime, the corrections to the EFT are parametrically controlled by the small string coupling constant and the large volume of the internal manifold in units of the string length.
  Since the non-perturbative effects breaking the PQ symmetry are suppressed as well,  stringy axions are typically light enough to be in the EFT.
  Then we expect that the behavior of  stringy axions in the EFT is dictated by the characteristic properties of string theory, and furthermore, quantum gravity.
 Indeed,  quantum gravity constraints on the dynamics of the EFT have drawn a lot of attention, but since only little is known about quantum gravity, they have been proposed as conjectures  under the name of `swampland program' \cite{Vafa:2005ui} (for reviews, see, e.g., \cite{Brennan:2017rbf, Palti:2019pca, vanBeest:2021lhn, Grana:2021zvf, Agmon:2022thq, VanRiet:2023pnx}).

 Quantum gravity constraints on the axion physics have been extensively studied in the context of the `weak gravity conjecture' (WGC) \cite{Arkani-Hamed:2006emk} (see \cite{Harlow:2022ich} for a review). 
It states that for consistency with quantum gravity, there must exist the U(1) charged particle whose gravitational interaction is weaker than the U(1) gauge interaction.
That is, given the gauge coupling $g$, the charge $q_e$ and the mass $m$  of such a particle satisfy the WGC bound $g q_e M_{\rm Pl}\gtrsim m$, where $M_{\rm Pl}$ is the $4$-dimensional Planck scale.
This can be supported by the argument that in the absence of the particle obeying the WGC bound, the extremal charged black hole cannot discharge through  the Hawking evaporation but decays into the charged remnant which is in trouble with the Bekenstein entropy bound \cite{Banks:2006mm}.
 The WGC can be extended to more generic Abelian gauge invariance in which the gauge field is given by the $p$-form $C_p$, by claiming the existence of a $(p-1)$-brane with the tension ${\cal T}_{p-1}$ and the charge $q_e$  satisfying $q_e M_{{\rm Pl}, d}^{\frac{d-2}{2}}\gtrsim {\cal T}_{p-1}$, where $M_{{\rm Pl}, d}$ is the $d$-dimensional Planck scale.
 Furthermore, we can treat the axion as the $0$-form gauge field by interpreting the PQ shift transformation  as the large gauge transformation.
 \footnote{While the gauge transformation typically indicates the local shift of the  $p$-form gauge field  given by $C_p\to C_p+d\lambda_{p-1}$, there also exists the large gauge transformation $C_p \to C_p+2\pi n \omega_p$, where $\omega_p$ is the harmonic $p$-form and $n$ is an integer, when the manifold contains the $p$-cycle.
For $p=0$, the local shift is not defined since there is no $(-1)$-form, but the large gauge transformation can be still defined.
Since $1$ is harmonic in any manifold ($d 1=0$ is trivial and $d*1=0$ because the rank of the volume form $*1$ is already maximal), the PQ shift transformation (more precisely, the residual one after the PQ symmetry breaking by the instanton effect) can be regarded as the large gauge transformation of the $0$-form gauge field, the axion.
For details, see, e.g., \cite{Reece:2023czb}.} 
In this case, the instanton action $S_{\rm inst}$ and the inverse of the axion decay constant $1/f$ play the roles of the tension and the charge, respectively, from which we obtain  the axion WGC bound $S_{\rm inst} f \lesssim M_{\rm Pl}$  \cite{Rudelius:2014wla, Rudelius:2015xta, Montero:2015ofa, Brown:2015iha, Hebecker:2015rya, Brown:2015lia, Heidenreich:2015wga}.

 On the other hand, many of swampland conjectures essentially rely on the `distance conjecture' \cite{Ooguri:2006in}, which states that the EFT beyond the asymptotic limit of the moduli space is invalidated by the descent of a tower of states, presumably either the string excitations or the Kaluza-Klein (KK) modes \cite{Lee:2019xtm, Lee:2019wij},  from UV.   
 Since the number of species in the EFT increases rapidly as the tower mass scale decreases, we expect that the cutoff scale would be lowered. 
 Indeed, it has been already noticed that the cutoff scale of quantum gravity is in fact lower than  $M_{\rm Pl}$ and determined by the number of  species in the EFT \cite{Veneziano:2001ah, Dvali:2007hz, Dvali:2007wp, Dvali:2009ks, Dvali:2010vm, Dvali:2012uq}.
 Above this cutoff scale   called  the species scale, the gravitational interaction becomes  strong since the number of species coupled to gravity is large enough to compensate the suppression by $M_{\rm Pl}^{-2}$. 
  Then the loop correction to the graviton propagator is much enhanced, so we can   say that quantum gravity effects are manifest at the energy scale above the species  scale.
 When a tower of states descends from UV as the distance conjecture predicts, the species scale decreases as well.
  
  In \cite{Reece:2024wrn}, it was pointed out that the connection between the cutoff scale and the number of species in the EFT also can be found in the axion physics without gravity.
  This comes from the fact that both the gravitational interaction  and the interaction through the $\theta F\wedge F$ term are non-renormalizable : these two are suppressed by the negative  powers of $M_{\rm Pl}$ and those of $\frac{8\pi^2}{g^2}f$, where $g$ is the gauge coupling, respectively.
Moreover, in the presence of the compact extra dimensional manifold, the axion also couples to the KK modes of the gauge field through the same form of the interaction term as $\theta F\wedge F$ with the same coupling $\frac{g^2}{8\pi^2 f}$, which will be explicitly shown in Sec. \ref{Sec:Straxion}.
 From this, in the same way as we did to define the species scale in quantum gravity,  we can define so-called the `axion species scale',  above which the suppression by the powers of the coupling $\frac{g^2}{8\pi^2 f}$ is compensated by the large number of   species, namely, the KK modes, or presumably, the string excitations.
 Now, the axion species scale is required to be lower than the species scale since gravity is assumed to be sufficiently weak that the gravitational interaction is not taken into account to obtain the axion species scale.
 This condition leads to the axion WGC-like bound, $\frac{8\pi^2}{g^2} f \lesssim M_{\rm Pl}$.
 \footnote{We refer the reader to \cite{Heidenreich:2017sim} which also claims that  the WGC can be obtained by requiring  the scale at which a gauge theory becomes strongly coupled  to be below the quantum gravity scale.
 This is motivated by the emergence proposal in which dynamics of gauge field  emerges from quantum gravity when a tower of charged particles (typically minimally couple  to gauge field) are integrated out. }
 For the non-Abelian gauge interaction, the factor $\frac{8\pi^2}{g^2}$ is identified with $S_{\rm inst}$, thus the bound is nothing more than the axion WGC bound.
 However, the interaction relevant to the axion WGC-like bound is not restricted to the non-Abelian gauge interaction, but extended to the Abelian gauge interaction and even to the string interaction : any interactions associated with towers of states can generate the bound.

  This article is devoted to study the axion species scale and the  axion WGC-like bound  discussed in the previous paragraph in detail in the context of  stringy axion, which is mainly addressed in Sec. \ref{Sec:AxSpSc}.
  For this purpose, we review in Sec. \ref{Sec:Straxion} the features of stringy axions needed for our discussion by considering the simple but typical examples, the model independent axion in the heterotic string theory and the axion from $C_4$ gauge field  in the type IIB string theory. 
  As we will see in Sec. \ref{Sec:AxSpSc}, in the simple stringy axion models, we can find the axion decay constant   saturating the axion WGC-like  bound, i.e.,  $\frac{8\pi^2}{g^2} f \sim M_{\rm Pl}$ is satisfied.
 In Sec. \ref{Sec:PQbreaking}, we investigate the implications of the (axion) species scale and the axion WGC-like bound on the PQ symmetry breaking.
  Since the PQ symmetry is a global symmetry, it must be broken by quantum gravity.
 The concept of the species scale implies that quantum gravity effects breaking the PQ symmetry become manifest at energy scale higher than the species scale, rather than $M_{\rm Pl}$. 
 At the same time, at this energy scale, the effects from the interaction between the axion and the gauge field through the $\theta F\wedge F$ term are strong as well, because the species scale is higher than the axion species scale.
  To see the consequences of considering the (axion) species scale explicitly, we discuss the reduction of the PQ charge of the axionic black hole \cite{Bowick:1988xh, Bowick:1989xg} (see also \cite{Hebecker:2017uix, Montero:2017yja} for recent discussions), which can take place through the interaction between the axionic black hole and the wormhole \cite{Giddings:1987cg} (for recent discussions, see, e.g., \cite{Hebecker:2016dsw, Hebecker:2018ofv}) or the  string  \cite{Hebecker:2017uix}.
  In both cases, the characteristic length scale of the PQ symmetry breaking is given by $\frac{1}{\sqrt {f M_{\rm Pl}}}$, and by comparing this with the (axion) species scale, we obtain the lower bound on the gauge coupling determined by the number of species.
 For the U(1) gauge interaction, this can be combined with so-called the Festina-Lente bound \cite{Montero:2019ekk}  (for phenomenological applications, see, e.g., \cite{Montero:2021otb, Lee:2021cor}), the upper bound on the gauge coupling depending on the IR scales.
 Then we obtain the condition that the U(1) charged particle mass must be larger than the Hubble parameter.

 \section{Review on Stringy axions}
 \label{Sec:Straxion}

 In this section, we review the basic properties of stringy axions with explicit examples, which are relevant to discussions in the following sections. 
 For the notations, we basically follow \cite{Blumenhagen:2013fgp}.
 When the string tension is given by $(2\pi \alpha')^{-1}$, the string length is defined as $\ell_s=2\pi\sqrt{\alpha'}$ and the  gravitational coupling in 10-dimensional supergravity is given by 
 \dis{\kappa_{10}^2=\frac{g_s^2}{4\pi}(4\pi \alpha')^4=\frac{g_s^2\ell_s^8}{4\pi},}
 where $g_s$ is the string coupling constant.

 In string theory, there are two ways of reducing the $p$-form gauge field $C_p$ to stringy axion under the compactification.
 First, we can obtain the axion by dualizing the $2$-form gauge field in four non-compact spacetime dimensions.
 This is called the model independent axion since its existence is irrelevant to the structure of the internal manifold.
 Besides this, another type of stringy axion comes from  $C_p$ with components along the directions of the  compactified internal manifold.
 Behaviors of stringy axions of this type in the moduli space are determined by the structure of the internal manifold.

 \subsection{Model independent axions}
 
 The massless spectrum of any string theory contains the $2$-form gauge field, either $B_2$ in the NS-NS sector or $C_2$ in the R-R sector.
The components of this $2$-form gauge field tangential to the $4$-dimensional non-compact spacetime can be dualized to the axion field, which is called the `model independent axion'.
 The essential features of the model independent axion can be found in the dualized Kalb-Ramond 2-form $B_2$ in the heterotic string theory.
The relevant terms of the heterotic string effective action are
  \dis{
  &-\frac{1}{2\kappa_{10}^2}\int \frac{\ell_s^4}{2} H_3 \wedge *_{10}H_3-\frac{1}{2g_{10}^2} \int {\rm Tr}(F  \wedge *_{10}F )
  \\
&=-\frac{2\pi}{g_s^2 \ell_s^4}\int \frac12 H_3 \wedge *_{10}H_3-\frac{1}{8\pi g_s^2 \ell_s^6} \int {\rm Tr}(F  \wedge *_{10}F ),\label{eq:hetS}}
  where the ratio between 10-dimensional gravitational ($\kappa_{10}$) and gauge ($g_{10}$) couplings, $g_{10}^2/\kappa_{10}^2=4/\alpha'$ is used.
  Here  $H_3$, the field strength of $B_2$, together with  the curvature 2-form $R$ and the gauge field strength $F$  satisfy the modified Bianchi identity
  \dis{dH_3=\frac{1}{16\pi^2}\Big[{\rm Tr}(R\wedge R)-{\rm Tr}(F \wedge F )\Big].}
  Since we are interested in the EFT of the axion in which the gravitational interaction is negligibly small, we omit $R\wedge R$ in the following discussion.
   The modified Bianchi identity can be imposed on the action by introducing the 6-form $B_6$ which plays the role of the Lagrange multiplier, and eventually, the dual field of $B_2$ : 
  \dis{-\frac{2\pi}{g_s^2 \ell_s^4}\int \frac12 H_3 \wedge *_{10}H_3-\frac{1}{8\pi g_s^2 \ell_s^6} \int {\rm Tr}(F  \wedge *_{10}F )+ \int B_6\wedge \Big[dH_3+\frac{1}{16\pi^2}{\rm Tr}(F \wedge F )\Big].}
 Integrating out $H_3$ gives the relation $H_7 \equiv dB_6=\frac{2\pi}{g_s^2\ell_s^4}*_{10}H_3$, from which  the action is written as
 \dis{-\frac{g_s^2\ell_s^4}{2\pi}\int \frac12 H_7\wedge *_{10}H_7 -\frac{1}{8\pi g_s^2 \ell_s^6} \int {\rm Tr}(F  \wedge *_{10} F )+\frac{1}{16\pi^2}\int B_6\wedge {\rm Tr}(F  \wedge F ).\label{eq:axionS}}
 Now   the heterotic string we are considering is compactified to 4-dimensions on some compact manifold $K_6$.
 We denote the 4-dimensional  non-compact manifold by $X_4$.
 In the KK reduction,  the 10-dimensional gauge field strength $F$ is expanded in terms of the eigenfunctions   $Y_{(a)}(y)$ of the differential operator  whose  eigenvalues determine the mass spectrum of the KK modes after compactification : 
  \dis{F = \sum_\alpha F^{(a)} (x) Y_{(a)}(y),}
  where the zero mode $F^{(0)} (x)$ ($a=0$) corresponds to   the massless 4-dimensional gauge field strength.
 The normalization condition for $Y_{(a)}(y)$ is given by $\int_{K_6} Y_{(a)}*_6Y_{(b)}={\rm Vol}(K_6)\delta_{ab}$.
 In addition, $B_6$ can be written as $\theta(x)\omega_6(x)$, where the $6$-form $\omega_6(y)=*_61/{\rm Vol}(K_6)$ is the normalized volume form of the internal manifold which is harmonic  (since $d1=0$ as well as $d(*_61)=0$).
 Then the first and the third term in \eqref{eq:axionS} become
 \footnote{Here we use the following relation between $q$-form $\eta_q(X_4)$ on $X_4$ and $r$-form $\xi_r(K_6)$ on $K_6$ :
 \dis{*_{10}(\eta_q(X_4)\wedge \xi_r(K_6))=(-1)^{qr}(*_4\eta_q(X_4)\wedge *_6\xi_r(K_6)).}
 For example, $*_{10}F =*_4 F \wedge *_61$, where $*_61$ is nothing more than the volume form of $K_6$, i.e., $\int_{K_6}*_61={\rm Vol}(K_6)$. }
 {\small
 \dis{&-\frac{g_s^2\ell_s^4}{2\pi}\Big(\int_{K_6}\frac{*_61\wedge 1}{{\rm Vol}(K_6)^2}\Big)\int_{X_4} \frac12 d\theta \wedge *_{4}d\theta +\frac{1}{16\pi^2}\sum_{a,b}\Big(\int_{K_6}\frac{*_61}{{\rm Vol}(K_6)}Y_{(a)}Y_{(b)}\Big)\int_{X_4} \theta  {\rm Tr}(F^{(a)}  \wedge F^{(b)} )
 \\
 &=-\int_{X_4}f^2 \frac12 d\theta \wedge *_{4}d\theta + \frac{1}{8\pi^2}\sum_a \int_{X_4}\frac12 \theta  {\rm Tr}(F^{(a)}  \wedge F^{(a)} ),\label{eq:MIaxion}}
 }
 where the squared axion decay constant is given by
 \dis{f^2=\frac{g_s^2\ell_s^4}{2\pi{\rm Vol}(K_6)}=\frac{g_s^2}{2\pi\ell_s^2{\cal V}}.\label{eq:hetf}}
 For the last equality we set ${\rm Vol}(K_6)={\cal V}\ell_s^6$.
 We also note that  the 4-dimensional Planck scale and the gauge coupling  (see \eqref{eq:hetS}) are given by 
 \dis{M_{\rm Pl}^2=\frac{4\pi{\cal V}}{g_s^2 \ell_s^2},\quad\quad
 g^2=\frac{4\pi g_s^2 }{{\cal V}}, \label{eq:hetg}}
 respectively, in terms of which the axion decay constant can be rewritten as
 \dis{f=\frac{g^2}{8\sqrt2 \pi^2} M_{\rm Pl}.\label{eq:Hetf}}
 The last term in \eqref{eq:MIaxion}  shows that the axion couples to the KK modes as well as the zero mode of the gauge field with the same coupling $\frac{g^2}{8\pi^2 f}$ (after canonical normalization of the axion and the gauge field).
 
 \subsection{Axions from $p$-form gauge field reduced on a $p$-cycle}
 
  On the other hand, string theory contains various $p$-form gauge fields $C_p$ where $p$ is not restricted to $2$.
  When reduced on $p$-cycles, they  give rise to axions in the 4-dimensional EFT.
  To be explicit, we note that the harmonic $p$-forms $\omega_{p, \alpha}$ ($\alpha=1,\cdots, b_p(K_6)$, where $b_p(K_6)$ is the $p$-th Betti number) on $K_6$ are the (unique) representatives of the cohomology classes, in terms of which $C_p$ is expanded as
  \dis{C_p (x,y)=\sum_{\alpha=1}^{b_p(K_6)} \frac{\theta^\alpha(x)}{2\pi} \omega_{p,\alpha}(y),}
  where the factor $\frac{1}{2\pi}$ is required to match the   coefficient of the $\theta F\wedge F$ term.  
 Defining the inner product of two harmonic forms by $\langle \omega_{p,\alpha}, \omega_{p,\beta}\rangle {\rm Vol}(K_6)=\int_{K_6} \omega_{p,\alpha}\wedge *_6 \omega_{p,\beta}$, and using the fact that $d\omega_{p,\alpha}=0$, one finds that the kinetic term for $C_p$ in the string frame becomes that for the axions $\theta^\alpha$,
\dis{-\frac{g_s^2}{2\kappa_{10}^2}\int \frac12 dC_p \wedge *dC_p=-\int_{X_4} f_{\alpha\beta}^2\frac12 d\theta^\alpha\wedge *_4 d\theta^\beta ,}  
where
\dis{f_{\alpha\beta}^2=\frac{g_s^2}{8\pi^2}M_{\rm Pl}^2\langle \omega_{p,\alpha}, \omega_{p,\beta}\rangle}
are the squared axion decay constants. 
When the compact internal manifold $K_6$ is given by Calabi-Yau 3-fold CY$_3$, the 4-dimensional EFT has ${\cal N}=2$ (${\cal N}=1$) supersymmetry for type II (type I and   heterotic) string theories.
The orientifold projection can reduce the supersymmetry from ${\cal N}=2$ to ${\cal N}=1$.   
In any case, the axion and the K\"ahler or  complex structure modulus belong to the same supersymmetry multiplet, thus can be treated as  a  pseudoscalar and scalar part of a complex scalar field, respectively.
Then information on the moduli space of CY$_3$ is encoded in the K\"ahler potential, in terms of which $f_{\alpha\beta}^2$ can be expressed as well.

As an example,  consider the 4-form gauge field $C_4$ in type IIB string theory, with the Standard Model gauge fields   living on D7-branes which wrap some of 4-cycles $\Sigma_4^\alpha$ ($\alpha=1, \cdots, h^{1,1}$).
The axions in this case can be obtained by the dimensional reduction of $C_4$ on $\Sigma_4^\alpha$.
Meanwhile, the K\"ahler form can be expanded as $\omega_{i \overline{j}}=\sum_i t_\alpha b^\alpha_{i\overline{j}}$, where $b^\alpha_{i\overline{j}}$ ($\alpha=1, \cdots, h^{1,1}$) form the basis of the harmonic $(1,1)$-forms, to which   the Poincar\'e duals to $\Sigma_4^\alpha$ belong.
Then   $t_\alpha$ correspond to the K\"ahler moduli, from which  the overall volume of the internal manifold (in units of the string length) is written as
\dis{{\cal V}=\frac16 \kappa^{\alpha\beta\gamma}t_\alpha t_\beta t_\gamma,}
where
\dis{\kappa^{\alpha\beta\gamma}=\int_{K_6} b^\alpha \wedge b^\beta \wedge b^\gamma}
are integer intersection numbers, and the volumes of 4-cycles are given by 
\dis{{\rm Vol}(\Sigma^\alpha_4)=\ell_s^4\tau^\alpha=\ell_s^4\frac{\partial {\cal V}}{\partial t_\alpha}=\frac{\ell_s^4}{2}\kappa^{\alpha\beta\gamma}t_\beta t_\gamma.}
Moreover, the K\"ahler potential is determined as
\dis{{\cal K} \equiv \frac{K}{M_{\rm Pl}^2}=-2  \log {\cal V}.}
For each $\alpha$, $t^\alpha$ and $\theta^\alpha$ correspond to the scalar and  pseudoscalar components of the chiral multiplet (hypermultiplet) of ${\cal N}=1$ (${\cal N}=2$) supersymmetry, thus  we can define complex scalar fields by $T^\alpha=\frac{1}{g_s}\tau^\alpha-i \frac{\theta^\alpha}{2\pi}$.
Imposing the PQ symmetry, the tree level K\"ahler potential is invariant under $\theta^\alpha \to \theta^\alpha+c^\alpha$, implying that it  depends only on $\tau^\alpha$, i.e., ${\cal K}={\cal K}(\frac{g_s}{2}(T^\alpha+{T^\alpha}^\dagger))$.
Then the kinetic term of $T^\alpha$ becomes
\dis{-\frac{\partial^2 K}{\partial T^\alpha \partial{T^\beta}^\dagger}\partial_\mu T^\alpha \partial^\mu {T^\beta}^\dagger =-\frac14 M_{\rm Pl}^2 \frac{\partial^2 {\cal K}}{\partial \tau^\alpha \partial{\tau^\beta}}\Big[\partial_\mu \tau^\alpha \partial^\mu \tau^\beta+ \frac{g_s^2}{4\pi^2}\partial_\mu \theta^\alpha \partial^\mu \theta^\beta\Big],}
from which the squared axion decay constants are given by
\dis{f_{\alpha\beta}^2=\frac{g_s^2}{8\pi^2}M_{\rm Pl}^2 \frac{\partial^2 {\cal K}}{\partial \tau^\alpha \partial{\tau^\beta}}.}

 Now the gauge fields on the D7-branes can be described by the Dirac-Born-Infeld (DBI) and  Chern-Simons (CS) actions. 
 From the DBI action, the U(1) gauge couplings are given by
 \dis{g_\alpha^2=\frac{g_{\rm D7}^2}{{\rm Vol}(\Sigma^\alpha_4)}=\frac{g_s(2\pi)^{p-2}(\alpha')^{\frac{p-3}{2}}\Big|_{p=7}}{{\rm Vol}(\Sigma^\alpha_4)}=\frac{2\pi g_s}{\tau^\alpha}.\label{eq:g2C4}}
 For the non-Abelian gauge couplings, the factor $2\pi$ in \eqref{eq:g2C4} is replaced by $4\pi$ when the gauge group generators are normalized as Tr$(T^a T^b)=\frac12\delta^{ab}$.
 Moreover, the CS action on the  D7-brane world-volume ${\cal W}_{\rm D7}$ contains
 \dis{\frac{\mu_7}{2}(2\pi \alpha')^2\int_{{\cal W}_{\rm D7}} C_4\wedge {\rm Tr}(F_\alpha\wedge F_\alpha)=\frac{1}{4\pi^2} \sum_{\alpha, a}\int_{X_4} \frac12  \theta^\alpha {\rm Tr}(F^{( a)}_\alpha\wedge F^{( a)}_\alpha).\label{eq:aFF4}}
 To reach the last expression, we use   $\mu_p=2\pi \ell_s^{-(p+1)}$ and   expand the field strength as $F_\alpha=\sum_{a}F^{( a)}_\alpha Y_{( a)}^\alpha$, where   $Y^\alpha_{(a)}$ are the eigenfunctions of the differential operator determining the masses of the KK modes.
 We also use   the  normalization condition $\int_{\Sigma_4^\alpha}\omega^\alpha Y_{(a)}^\alpha Y_{( b)}^\alpha=\ell_s^4 \delta_{ab}$.
 \footnote{Comparing with the case of the model independent axion, $\omega^\alpha$ can be regarded as the volume form of $\Sigma_4^\alpha$ divided by ${\rm Vol}(\Sigma_4^\alpha)$.
 Then the normalization condition is   given by $\int_{\Sigma_4}Y_{(a)}^\alpha *_4Y_{(b)}^\alpha=\ell_s^4 {\rm Vol}(\Sigma_4^\alpha)\delta_{ab}$, just like that for the model independent axion.
 The factor $\ell_s^4$ is for the dimensional consistency, which corresponds to the rescaling of  $H_3$ by $\ell_s^2 H_3$  in \eqref{eq:hetS}. }
 Then just like the case of the model independent axion, the axion $\theta^\alpha$ couples to the KK modes as well as the zero mode  of the gauge field with the same coupling $\frac{g_\alpha^2}{8\pi^2 f}$.
 From the discussion so far, we find that  the axion decay constant may be rewritten as
 \dis{f_{\alpha\beta}=\frac{g_\alpha g_\beta \sqrt{\tau^\alpha\tau^\beta}}{4\sqrt2 \pi^2}M_{\rm Pl} \Big(\frac{\partial^2 {\cal K}}{\partial \tau^\alpha \partial{\tau^\beta}}\Big)^{1/2},\label{eq:C4f}}
  because the combination $g_\alpha^2\tau^\alpha$ is just given by $2\pi g_s$, which is independent of the particular choice of $\Sigma_4^\alpha$.
  For the non-Abelian gauge interaction, the factor $\frac12$ is multiplied in addition.

 \section{Species scale and axion species scale}
 \label{Sec:AxSpSc}
 
  We now investigate the axion species scale determined by the number of  species in the EFT, in the presence of the non-renormalizable interaction term $\theta F\wedge F$.
    For comparison, we begin our discussion with the review on the species scale, the quantum gravity cutoff determined in the same way but   the relevant interaction is   the  gravitational interaction which is also non-renormalizable.
 
 \subsection{Species scale in quantum gravity}
 
The species scale above which gravity is no longer weakly coupled can be defined in several ways, giving the same result.
Here we pursue the idea that defines the species scale as a scale at which the 1-loop correction to the graviton propagator becomes ${\cal O}(1)$, signalling the breakdown of the perturbativity \cite{Dvali:2007wp} (see also \cite{Castellano:2022bvr}). 
To see this concretely, we recall that the 1-loop corrected graviton propagator in 4-dimensional spacetime   is given by \cite{Calmet:2017omb}
\dis{i D_{\alpha\beta,\mu\nu}(p^2)=i\big(L_{\alpha\mu}L_{\beta\nu}+L_{\alpha\nu}L_{\beta\mu}-L_{\alpha\beta}L_{\mu\nu}\big) G_g(p^2),}
where
\dis{L_{\mu\nu}=\eta_{\mu\nu}-\frac{p_\mu p_\nu}{p^2}} 
and
\dis{G_g^{-1}(p^2)=2p^2\Big[1-\overline{c}_L N \frac{p^2}{M_{\rm Pl}^2}\log\Big(-\frac{p^2}{\mu^2}\Big)\Big].\label{eq:gravprop}}
In \eqref{eq:gravprop}, $\overline{c}_L$ is a loop factor  and $N$ is the number of   species in the EFT contributing to the loop.  
From this, one finds that the 1-loop correction given by the second term in the bracket in \eqref{eq:gravprop} is larger than ${\cal O}(1)$ when $-p^2$ is larger than the squared species scale given by
\dis{\Lambda_{\rm sp}^2\simeq -2 \frac{M_{\rm Pl}^2}{N}\Big[W_{-1}\Big(-\frac{2}{N}\frac{M_{\rm Pl}^2}{\mu^2}\Big)\Big]^{-1}\simeq \frac{M_{\rm Pl}^2}{N \log N} \sim \frac{M_{\rm Pl}^2}{N},\label{eq:grav1}}
where $W_{-1}(x)$ is the $(-1)$-branch of Lambert W function, which is approximated by $\log(-x)-\log(-\log(-x))+\cdots$ in the limit $x\to 0^-$.
Then $N$, which will be denoted by $N_{\rm sp}$ in the discussion below,  counts the number of  species   below $\Lambda_{\rm sp}$ (in other words, the number of   species in the EFT with the cutoff $\Lambda_{\rm sp}$).

Now, in order to combine the concept of the species scale with the distance conjecture, consider a tower of states contributing to the loop, the spectrum of which can be written as
\dis{m_n = n^{1/p} m_t,}
where the positive integers $n$ and $p$ represent the step of the state within the tower and the number of particles with  identical mass gap, respectively \cite{Castellano:2021mmx}.
For the  KK tower ($m_t = m_{\rm KK}$), $p$ is identified with the number of towers with the identical value of $m_{\rm KK}$ and $N_{\rm sp}$ is estimated as $(\Lambda_{\rm sp}/m_{\rm KK})^p$. 
For the string tower ($m_t=m_{\rm string}=\ell_s^{-1}$), whereas $m_n \sim \sqrt{n} m_{\rm string}$, the degeneracy grows exponentially as $d_n \sim e^{\sqrt{n}}$ \cite{Kani:1989im}, which corresponds to the limit $p\to \infty$.
Therefore, we obtain the generic relation $N_{\rm sp} \sim (\Lambda_{\rm sp}/m_{t})^p$.
Combining this with \eqref{eq:grav1}, one finds that
\dis{&\Lambda_{\rm sp} \sim  M_{\rm Pl}^{\frac{2}{p+2}}m_{t}^{\frac{p}{p+2}},\quad\quad
N_{\rm sp}\sim   \Big(\frac{M_{\rm Pl}}{m_{t}}\Big)^{\frac{2p}{p+2}}.\label{eq:spG}} 
In the presence of the compact extra-dimensional manifold $\Gamma_p$, the KK towers appear, in which case $\Lambda_{\rm sp}$ and $N_{\rm sp}$ are estimated as
\dis{\Lambda_{\rm sp} \sim \Big(\frac{M_{\rm Pl}^2}{{\rm Vol}(\Gamma_p)}\Big)^\frac{1}{p+2},\quad\quad  N_{\rm sp}\sim M_{\rm Pl}^{\frac{2p}{p+2}}({\rm Vol}(\Gamma_p))^{\frac{2}{p+2}},}
  implying that $\Lambda_{\rm sp}$ is given by the $(4+p)$-dimensional Planck scale.
 On the other hand, for the string tower,
\dis{&\Lambda_{\rm sp} \sim   m_{\rm string} =\frac{1}{\ell_s},\quad\quad
N_{\rm sp}\sim   \Big(\frac{M_{\rm Pl}}{m_{\rm string}}\Big)^{2}=(M_{\rm Pl}\ell_s)^2,}
are satisfied.

 We close this subsection by discussing  the species scale in the presence of multiple towers with different tower mass scales \cite{Castellano:2021mmx, Castellano:2022bvr, Seo:2023xsb}.
 When the mass spectra of towers are given by
 \dis{m_{n_1}=n_1^{1/p_1}m_{t_1},\quad m_{n_2}=n_2^{1/p_2}m_{t_2}, \quad \cdots,}
 we can  find the species scales associated with each tower,
 \dis{\Lambda_{{\rm sp}, i} \sim  M_{\rm Pl}^{\frac{2}{p_i+2}}m_{t_i}^{\frac{p_i}{p_i+2}},\quad
N_{{\rm sp}, i}\sim   \Big(\frac{M_{\rm Pl}}{m_{t_i}}\Big)^{\frac{2p_i}{p_i+2}},\quad i=1, 2, \cdots.\label{eq:assSp}}
Then the physically meaningful species scale is $\Lambda_{\rm sp}={\rm min}\{\Lambda_{{\rm sp}, i}\}$.
 On the other hand, we can easily find the state carrying  several tower excitation numbers simultaneously, such that it is labelled by $(n_1, n_2, \cdots )$ and the squared mass is given by $m_{(n_1, n_2,\cdots)}=m_{n_1}^{2/p_1}m_{t_1}^2+m_{n_2}^{2/p_2}m_{t_2}^2+\cdots$.
 As an example, we may consider the KK mode in the presence of  different KK mass scales   arising from the different sizes of the cycles in the internal manifold.
 In this case, the number of   species   below $\Lambda_{\rm sp}$ is given by $N_{\rm sp}=N_1N_2\cdots$, where
 \dis{N_1=\Big(\frac{\Lambda_{\rm sp}}{m_{t_1}}\Big)^{p_1},\quad N_2=\Big(\frac{\Lambda_{\rm sp}}{m_{t_2}}\Big)^{p_2},\quad\cdots.\label{eq:each}}
 Then from $\Lambda_{\rm sp}=M_{\rm Pl}/\sqrt{N_{\rm sp}}$, we obtain
 \dis{\Lambda_{\rm sp}=M_{\rm Pl}^{\frac{2}{p+2}} \prod_i m_{t_i}^{\frac{p_i}{p+2}}, \quad
 N_{\rm sp}=\frac{M_{\rm Pl}^{\frac{2p}{p+2}}}{\prod_i m_{t_i}^{\frac{2p_i}{p+2}}  },\label{eq:spfin}}
 where $p=p_1+p_2+\cdots$.

\subsection{Axion  species scale }
\label{sec:AxSpSc}

Meanwhile, \cite{Reece:2024wrn} observed that the UV cutoff similar to the species scale can be obtained   from the 1-loop correction to the axion propagator to which the KK modes of the gauge field contribute.
 Discussion in Sec. \ref{Sec:Straxion} indicates that the coupling between the canonically normalized fields $f\theta$ and $F/g$ through the $\theta F\wedge F$ term is given by $\frac{g^2}{8\pi^2 f}$.
From this, the 1-loop corrected axion propagator is estimated as
\dis{G_a^{-1}(p^2)=p^2\Big[1+c_L\Big( \frac{g^2}{8\pi^2 f}\Big)^2N p^2 \log\Big(-\frac{p^2}{\mu^2}\Big)\Big],\label{eq:Ga}}
where $N$ is the number of KK modes with mass below $|q|$.
 Using the same argument as in the case of the graviton propagator, we find that the perturbativity breaks down for $-q^2$ larger than the squared `axion species scale',
\dis{\Lambda_{{\rm sp}, a}^2 \sim \Big(\frac{8\pi^2 f}{g^2}\Big)^2\frac{1}{N_{{\rm sp}, a}},\label{eq:Lspa1}}
where we denote   the number of KK modes below $\Lambda_{{\rm sp}, a}$ by $N_{{\rm sp}, a}$, which is given by $(\Lambda_{{\rm sp}, a}/m_{\rm KK})^r$ when the axion couples to $r$ KK towers of the gauge field with the same mass scale.
\footnote{Here $r$ does not necessarily coincide with the exponent $p$ in $N_{\rm sp}=(\Lambda_{{\rm sp}}/m_{\rm KK})^p$ since in general not all towers couple to the axion. 
Of course,  $r\leq p$ must be satisfied.
}
Then we obtain
\dis{&\Lambda_{{\rm sp},a} \sim \Big(\frac{8\pi^2}{g^2}f\Big)^\frac{2}{r+2}\Big(\frac{1}{{\rm Vol}(\Gamma_p)}\Big)^\frac{1}{r+2}\sim \Big(\frac{8\pi^2}{g^2}f\Big)^\frac{2}{r+2} m_{\rm KK}^{\frac{r}{r+2}},
\\
&N_{{\rm sp},a}\sim \Big(\frac{8\pi^2}{g^2}f\Big)^{\frac{2r}{r+2}}({\rm Vol}(\Gamma_p))^{\frac{2}{r+2}}\sim \Big(\frac{8\pi^2}{g^2}\frac{f}{m_{\rm KK}}\Big)^{\frac{2r}{r+2}},\label{eq:spa}}
which correspond to the replacement of $M_{\rm Pl}$ in $\Lambda_{\rm sp}$ and $N_{\rm sp}$ by $\frac{8\pi^2}{g^2}f$.
Moreover, the above result may be extended to the string tower, not   restricted to the KK tower.
 To see this, we recall that stringy axions are the massless modes of the closed string, and comparing with the tree level one, the string 1-loop amplitude   contains a factor $g_s^2$ in addition.
Since the axion decay constant $f$ comes from the canonical normalization of the axion,  the external state of the propagator, the dimensional consistency implies that the string tower also leads to the same expression  for $\Lambda_{{\rm sp},a}$ and $N_{{\rm sp},a}$ as \eqref{eq:spa}, in which $g^2$ and $m_{\rm KK}$ are given by $g_s$ and $m_s=\ell_s^{-1}$, respectively, and the $r\to \infty$ limit is taken :
 \dis{&\Lambda_{{\rm sp},a}  \sim  m_{\rm string}=\frac{1}{\ell_s},
\quad\quad
N_{{\rm sp},a}\sim \Big(\frac{8\pi^2}{g_s}\frac{f}{m_{\rm string}}\Big)^2 = \Big(\frac{8\pi^2}{g_s}f \ell_s \Big)^2.}
 Then the generic relations satisfied by both the string and KK towers,
 \dis{&\Lambda_{{\rm sp},a}\sim \Big(\frac{8\pi^2}{g^2}f\Big)^\frac{2}{r+2} m_{t, a}^{\frac{r}{r+2}},
\quad\quad
 N_{{\rm sp},a} \sim \Big(\frac{8\pi^2}{g^2}\frac{f}{m_{t, a}}\Big)^{\frac{2r}{r+2}},\label{eq:spA}}
 can be written.
 Here we denote the mass scale of the tower which couples to the axion by    $m_{t, a}$.
  It of course belongs to $\{m_t\}$ that defines $\Lambda_{\rm sp}$. 
  On the other hand, the axion can couple to  multiple towers.
  When the state which couples to the axion carries  several tower excitation numbers simultaneously,
   $\Lambda_{{\rm sp}, a}$ and $N_{{\rm sp}, a}$ can be written in the same way as $\Lambda_{\rm sp}$ and $N_{\rm sp}$ in \eqref{eq:spfin} : 
 \dis{\Lambda_{{\rm sp}, a}=\Big(\frac{8\pi^2}{g^2}f\Big)^\frac{2}{r+2} \prod_i m_{t_i, a}^{\frac{r_i}{r+2}},\quad \quad
 N_{{\rm sp}, a}=\prod_i \Big(\frac{\Lambda_{{\rm sp}, a}}{m_{t_i, a}}\Big)^{r_i}=\frac{\Big(\frac{8\pi^2}{g^2}f\Big)^{\frac{2r}{r+2}}}{\prod_i m_{t_i, a}^{\frac{2r_i}{r+2}}  }, \label{eq:spAmulti}}
 where $r=r_1+r_2+\cdots$.
 We note that even though  denoted differently, $\{ r_i\}$ is a subset of $\{ p_i\}$, since the tower which couples to the axion  couples to gravity as well.
 Moreover, unlike the gravitational interaction, the coupling of the axion to the states in the loop is not universal, but given by the gauge  (or string) coupling.
 In this case, we need to find out the axion species scales associated with different couplings.
 Then the axion species scale is identified with the minimum of them.

 Now, recall that  in obtaining the axion species scale, any gravitational effect is not taken into account.
 That is,  gravity is presumed to be  weakly coupled at energy scale lower than $\Lambda_{{\rm sp},a}$, thus $\Lambda_{{\rm sp},a} \leq \Lambda_{\rm sp}$ is satisfied.
 We will quantify this condition in two ways.
 First, when every state carries the single tower excitation number only, we can find the (gravitational) species scales associated with  each of towers which couple to the axion.
 They form a subset of the set of the species scales  in \eqref{eq:assSp} :
 \dis{\Lambda'_{{\rm sp}, i, a} \sim  M_{\rm Pl}^{\frac{2}{r_i+2}}m_{t_i, a}^{\frac{r_i}{r_i+2}},\quad
N_{{\rm sp}, i, a}'\sim   \Big(\frac{M_{\rm Pl}}{m_{t_i, a}}\Big)^{\frac{2r_i}{r_i+2}},\quad i=1, 2, \cdots,}
and their minimum value min$\{\Lambda'_{{\rm sp}, i, a}\}$  can be interpreted as the species scale associate with towers which couple to the axion.
 \footnote{Here the prime is used to emphasize that, whereas we consider towers that couple to the axion, the UV scale is taken to be $M_{\rm Pl}$, not $\frac{8\pi^2}{g^2}f$.} 
 In addition, we can also define the axion species scales associated with towers as
  \dis{\Lambda_{{\rm sp}, i, a} \sim  \Big(\frac{8\pi^2}{g^2}f\Big)^{\frac{2}{r_i+2}}m_{t_i, a}^{\frac{r_i}{r_i+2}},\quad
N_{{\rm sp}, i, a}\sim   \Big(\frac{\frac{8\pi^2}{g^2}f}{m_{t_i, a}}\Big)^{\frac{2r_i}{r_i+2}},\quad i=1, 2, \cdots,}
and take their minimum value.
In fact, the tower mass scale associated with min$\{\Lambda_{{\rm sp}, i, a}\}$ is nothing more than that associated with min$\{\Lambda'_{{\rm sp}, i, a}\}$.
Demanding min$\{\Lambda_{{\rm sp}, i, a}\}\lesssim $min$\{\Lambda'_{{\rm sp}, i, a}\}$, we obtain the bound $\frac{8\pi^2}{g^2}f \lesssim M_{\rm Pl}$.
Whereas it is quite attractive as it resembles the axion WGC bound, we need to note that min$\{\Lambda'_{{\rm sp}, i, a}\}$ is not necessarily identical to the true species scale, min$\{\Lambda_{{\rm sp}, i}\}$.
This is because to obtain min$\{\Lambda'_{{\rm sp}, i, a}\}$, we just consider the towers that couple to the graviton and the axion simultaneously.
In some sense, this is similar in spirit to the WGC bound : it also  considers the particle which couples to gravity and the Abelian gauge interaction simultaneously.  
 
 On the other hand, when the state carries several tower excitation numbers simultaneously, the condition that $\Lambda_{\rm sp, a}$ in \eqref{eq:spAmulti} is smaller than $\Lambda_{\rm sp}$ reads
 \dis{\frac{8\pi^2}{g^2}\frac{f}{M_{\rm Pl}} \lesssim \frac{\Lambda_{\rm sp}}{M_{\rm Pl}}\prod_i \Big(\frac{\Lambda_{\rm sp}}{m_{t_i, a}}\Big)^{r_i/2}=\frac{1}{\sqrt{N_{\rm sp}}}\prod_i \Big(\frac{\Lambda_{\rm sp}}{m_{t_i, a}}\Big)^{r_i/2}.\label{eq:bound1}}
 To interpret the RHS, we need to note from \eqref{eq:each} that $N'_{i, a}\equiv (\Lambda_{\rm sp}/m_{t_i, a})^{r_i}$ is the contribution of the specific tower  to $\Lambda_{\rm sp}$.
 This tower couples to the axion and characterized by the mass scale $m_{t_i, a}$ and the exponent $r_i$.
 While $N_{\rm sp}$ is given by the product of $N_i$s from all the towers in the EFT, the product 
 \dis{N'_{{\rm sp}, a} \equiv \prod_i N'_{i, a} = \prod_i \Big(\frac{\Lambda_{\rm sp}}{m_{t_i, a}}\Big)^{r_i}}
 is the total contribution  of the towers which couple to the axion to $N_{\rm sp}$.
 In other words, $\{ N'_{i, a}\} \subset \{N_i \}$ hence $N_{\rm sp}$ is given by the product of $N'_{{\rm sp}, a}$ and $N_i$s for the towers which do not couple to the axion.
 Therefore, the RHS of \eqref{eq:bound1} given by $\sqrt{N'_{{\rm sp}, a}/N_{\rm sp}}$ is always smaller than $1$, resulting in the axion WGC-like bound $\frac{8\pi^2}{g^2}f\lesssim M_{\rm Pl}$.

 In the simplest but unrealistic case in which all the towers couple to the axion, from $N_{\rm sp}=N'_{{\rm sp},a}$, the  axion WGC-like bound is easily obtained.
 However, since gravity can couple to towers to which the axion does not couple,  it is typical that $N'_{{\rm sp},a}$ is much smaller than $N_{\rm sp}$, implying that the axion WGC-like bound is not saturated.
 This contrasts to the value of $f$ obtained from the concrete string theory model : as we will see, we can easily find the case in which the bound is saturated, such that  $\frac{8\pi^2}{g^2}f = M_{\rm Pl}$ is satisfied.
 In fact, this paradoxical situation can be circumvented if one of $r_i$, say, $r_1$, which indeed belongs to $\{p_i\}$ is infinitely large.
  In this case, we can infer from \eqref{eq:spfin}  that $\Lambda_{\rm sp}\simeq m_{t_1, a}$ and $N_{\rm sp}\simeq (M_{\rm Pl}/m_{t_1, a})^2$, and from \eqref{eq:spAmulti} that $\Lambda_{{\rm sp}, a}\simeq m_{t_1, a}$ and $N_{{\rm sp}, a}\simeq (\frac{8\pi^2}{g^2}f/m_{t_1, a})^2$, respectively.
 From $\Lambda_{\rm sp}\simeq \Lambda_{{\rm sp}, a} \simeq  m_{t_1, a}$, the inequality \eqref{eq:bound1} can be approximated by 
 \dis{\frac{8\pi^2}{g^2}\frac{f}{M_{\rm Pl}} \lesssim  \frac{1}{\sqrt{N_{\rm sp}}}\prod_i \Big(\frac{\Lambda_{\rm sp}}{m_{t_i, a}}\Big)^{r_i/2} \simeq 
 \frac{1}{\sqrt{N_{\rm sp}}}\prod_i \Big(\frac{\Lambda_{{\rm sp}, a}}{m_{t_i, a}}\Big)^{r_i/2}=\sqrt{\frac{N_{{\rm sp}, a}}{N_{\rm sp}}}, }
 where for the last expression we used \eqref{eq:spAmulti}, which is just given by $\frac{8\pi^2}{g^2}\frac{f}{M_{\rm Pl}}$.
 That is,  the inequality is saturated.
 Since the RHS of this inequality,  an approximation of $\sqrt{N'_{{\rm sp}, a}/N_{\rm sp}}$, is smaller than $1$, we obtain the axion WGC-like bound $\frac{8\pi^2}{g^2}f\lesssim M_{\rm Pl}$ again.
The point here is that the ratio $\sqrt{N'_{{\rm sp}, a}/N_{\rm sp}}$ is identified with the ratio $\frac{8\pi^2}{g^2}\frac{f}{M_{\rm Pl}}$, and there is no any other constraints on this value other than it is similar to or smaller than $1$.
 Indeed, in the presence of the tower with $r_i\to\infty$, contributions from other towers to both $N'_{{\rm sp}, a}$ and $N_{\rm sp}$ are not important.
 Moreover, \eqref{eq:spAmulti} tells us that in this case the contribution of the highest mass scale, namely, $\frac{8\pi^2}{g^2}f$, to $\Lambda_{{\rm sp}, a}$ is suppressed.
 Then even if the axion couples to a number of towers of different couplings, it may not be difficult for the axion species scale associated with infinitely large $r_i$ to be the minimum.

 In summary,  the condition that the axion species scale is smaller than the species scale leads to the axion  WGC-like bound given by 
  \dis{\frac{8\pi^2}{g^2}f \lesssim M_{\rm Pl}.\label{eq:aWGC}}
For the non-Abelian gauge field, the instanton action is given by $\frac{8\pi^2}{g^2}$, hence \eqref{eq:aWGC} is nothing more than the axion WGC bound $ S_{\rm inst} f \lesssim M_{\rm Pl}$. 
 On the other hand, for the Abelian gauge field, the factor  $\frac{8\pi^2}{g^2}$ cannot be interpreted as the instanton action, but still we can impose the bound \eqref{eq:aWGC}   since the factor in our argument comes from the coupling between the axion and (the KK modes of) the gauge field, which is irrelevant to the instanton if we restrict attention to the gauge field and the U(1) charged particles only.
 \footnote{If we introduce the Euclidean D-branes as extended objects, the instanton action $\sim \frac{1}{g_s}$(volume of the brane) can be estimated as $\frac{1}{g^2}$ (see \eqref{eq:g2C4}), where the gauge group associated with the gauge coupling $g$ may be Abelian.
 Moreover, in the presence of the dyon, the instanton can be generated by its loop contributions even in the case of the Abelian gauge interaction \cite{Fan:2021ntg} (we thank an anonymous referee for pointing out this). 
  }
 Therefore, in the axion WGC-like bound, the gauge group is not restricted to the non-Abelian, but can be extended to the Abelain as well.
 Another remarkable feature is that, when the state carries several tower excitation numbers simultaneously, saturation of the axion WGC-like bound is allowed in the presence of the tower of infinitely large $r_i$ value.
 Since a typical example of the tower of infinitely large $r_i$ value is the string tower, it may imply the close connection between the axion WGC-like bound (and perhaps the WGC) and  string theory.

 In the simplest string theory model, the axion WGC-like bound is typically saturated, which can be found in the examples in Sec. \ref{Sec:Straxion} as well \cite{Svrcek:2006yi}.
For the model independent axion in the heterotic string theory, the relation \eqref{eq:Hetf}  can be rewritten as
\dis{\frac{8\pi^2 }{g^2}f = \frac{1}{\sqrt2} M_{\rm Pl},} 
so the axion WGC-like bound is saturated.
On the other hand, for the axions from $C_4$, as can be found in  \eqref{eq:aFF4}, the gauge field $F_\alpha$   contributes to the loop correction to the  axion $\theta^\beta$ propagator  provided $\alpha=\beta$ , i.e., the axion and the gauge fields are associated with the same 4-cycle.
Denoting the associated gauge coupling by $g_\alpha$, from  \eqref{eq:C4f}  we obtain  
\dis{\frac{8\pi^2 }{g_\alpha^2}f_{\alpha\alpha} =  \sqrt2 M_{\rm Pl}\tau^\alpha \Big(\frac{\partial^2 {\cal K}}{\partial {\tau^\alpha}^2  }\Big)^{1/2},}
where the indices are not contracted.
Even though $\tau^\alpha>1$ is required for the validity of the EFT, $\partial_{\tau^\alpha}^2{\cal K}$ is typically suppressed by some power of the overall volume ${\cal V}$, which prevents  $\frac{8\pi^2}{g^2}f$ from being   enhanced compared to $M_{\rm Pl}$.
In order to see how this works, we consider the concrete model, an orientifold of CY$_3$ given by the degree 18 hypersurface in $\mathbb{P}^4_{[1,1,1,6,9]}$.
This has been studied in \cite{Denef:2004dm} (see also \cite{Candelas:1994hw} for an earlier discussion) and used to realize the large volume scenario \cite{Balasubramanian:2005zx}.
In this model, there are  two 4-cycles, one (denoted by $\tau_5$) controls the size of the overall volume, while another (denoted by  $\tau_4$) determines the size of small hole in the manifold.
The volume is given by
\dis{{\cal V}=\frac{1}{9\sqrt2}(\tau_5^{3/2}-\tau_4^{3/2}),\label{eq:LVS}}
and the moduli stabilization results in $\tau_5\sim {\cal V}^{2/3}$ and $\tau_4\sim \log{\cal V}$. 
Therefore, we obtain
\dis{&\frac{8\pi^2}{g_5^2}\frac{f_{55}}{\sqrt2 M_{\rm Pl}}= \tau_5 \Big(\frac{\partial^2 {\cal K}}{\partial {\tau_5}^2  }\Big)^{1/2}=\tau_5\Big[-\frac{\sqrt2}{12}\frac{1}{{\cal V}\tau_5^{1/2}}+\frac{1}{36}\frac{\tau_5}{{\cal V}^2}\Big]^{1/2}
\sim {\cal O}(1),
\\
&\frac{8\pi^2}{g_4^2}\frac{f_{44}}{\sqrt2 M_{\rm Pl}}=\tau_4 \Big(\frac{\partial^2 {\cal K}}{\partial {\tau_4}^2  }\Big)^{1/2} =\tau_4 \Big[\frac{\sqrt2}{12}\frac{1}{{\cal V}\tau_4^{1/2}}+\frac{1}{36}\frac{\tau_4}{{\cal V}^2}\Big]^{1/2} \sim \frac{(\log{\cal V})^{3/4}}{{\cal V}^{1/2}},\label{eq:fcheese}} 
which shows that whereas the axion WGC-like bound for the axion associated with $\tau_5$ is saturated,   that for the axion associated with $\tau_4$ is volume suppressed ($\sim 1/\sqrt{\cal V}$).
 We also note   from \eqref{eq:C4f} and the relation $t_\alpha=\partial_{\tau^\alpha}\partial_{\tau^\beta}{\cal K}\tau^\beta{\cal V}$ (for a proof, see, e.g., App. A of \cite{Reece:2024wrn})  that the inequality
 \dis{\sum_{\alpha, \beta} f_{\alpha\beta}^2& \leq \frac{({\rm max}(g_\alpha^2))^2}{32 \pi^4} \sum_{\alpha, \beta}  \tau^\alpha\tau^\beta\frac{\partial^2 {\cal K}}{\partial \tau^\alpha \partial{\tau^\beta}}   M_{\rm Pl}^2=\frac{({\rm max}(g_\alpha^2))^2}{32 \pi^4}\sum_\alpha \frac{\tau^\alpha t_\alpha}{\cal V}  M_{\rm Pl}^2
 \\
 &=\Big[\frac{{\rm max}(g_\alpha^2)}{4\sqrt2 \pi^2}\times \sqrt3 M_{\rm Pl}\Big]^2,}
 is satisfied,  which can be another way to express the axion WGC-like bound.
In our model, in addition to \eqref{eq:fcheese}, 
\dis{ \tau_5\tau_4 \frac{\partial^2 {\cal K}}{\partial {\tau_5}\partial {\tau_4}} =\tau_5\tau_4\Big(-\frac{1}{36}\frac{\tau_5^{1/2}\tau_4^{1/2}}{{\cal V}^2}\Big)  \sim \frac{(\log{\cal V})^{3/4}}{{\cal V}^{1/2}},\label{eq:fcheese2}}
is satisfied, which contributes to the bound above.

\section{Peccei-Quinn symmetry breaking }
\label{Sec:PQbreaking}

Since the PQ symmetry is a global symmetry, it is broken  by quantum gravity effects.
In discussing the PQ symmetry breaking of this type, it has been typically assumed that quantum gravity effects become strong at the energy scale above $M_{\rm Pl}$.
However, the concept of the species scale suggests that $M_{\rm Pl}$ in the assumption needs to be replaced by $\Lambda_{\rm sp}$.
Moreover, 
we can infer from our previous discussion that the effects of the $\theta F\wedge F$ term is no longer negligible at the energy scale above the axion species scale.
Since the axion species scale $\Lambda_{{\rm sp}, a}$    is lower than $\Lambda_{{\rm sp} }$,  when the PQ symmetry breaking is manifest,  both quantum gravity and the $\theta F\wedge F$ term effects must be strong.
 In this section, we investigate the PQ symmetry breaking  in light of the (axion) species scale and the axion WGC-like bound, by considering   the PQ charge reduction of the axionic black hole, the black hole carrying the PQ charge.

 The axionic black hole is a static black hole solution with mass $M$ and  PQ charge $q$ given by \cite{Bowick:1988xh, Bowick:1989xg} 
 \dis{&ds^2=-\Big(1-\frac{2 GM}{r}\Big)dt^2+\frac{dr^2}{\Big(1-\frac{2 GM}{r}\Big)}+r^2d\Omega_2^2,
 \\
 &(B_2)_{\mu\nu}=\frac{q}{4\pi }d\Omega_2 ,\label{eq:axBH}}
 where  $B_2$ is the dual field of the axion in four spacetime dimensions,  i.e., $H_3=dB_2=f^2 *_4d\theta$, which is familiar from the model independent axion.
 The PQ charge $q=\int_{S^2} B_2$  is conserved as $S^2$ corresponds to the homology class of the event horizon in the spacetime with the singularity excised.
 It means that $q$ is a topological charge, rather than the charge associated with a local symmetry, thus it is locally unobservable.
 This is evident from the fact that $B_2$ in the  solution \eqref{eq:axBH} is a pure gauge, which also implies that $B_2$ does not carry energy and the mass $M$ is independent of the size of $q$ : the axionic black hole of given mass $M$ can have any value of $q$.
 
 This property of the axionic black hole raises a subtle issue at the quantum level.
 Since the PQ charged particles couple to $H_3$, not $B_2$,  the vanishing value of $H_3$ in the solution \eqref{eq:axBH} means that the PQ charge of the black hole is not radiated away by the Hawking evaporation.
 Then the black hole with a large $q$ cannot  totally evaporate for the following reason : if the axionic black hole is gone, spacetime does not have a singularity any longer, then the large amount of $q$ which was stored in the singularity  must be carried by $B_2$ with the non-vanishing $H_3$ satisfying $\int_V H_3 =q$, where the  volume $V$ is slightly larger than that of the smallest black hole described by the EFT. 
 Since  the non-vanishing $H_3$  carries the energy $\frac{1}{f^2}\int_V |H_3|^2$ as well, the  large value of $q$ (hence $H_3$), or equivalently, the small value of $f$ leads to the energy much larger than the smallest black hole mass, which is not consistent with the conservation of   energy.  
 While this seems to imply that the final state of the axionic black hole must be a remnant,  the existence of the remnant is quite controversial as its entropy may violate the Bekenstein bound \cite{Susskind:1995da}.
 
 In order to resolve this paradox, there must be the process that reduces the PQ charge of the black hole, which allows  the axionic black hole to evaporate completely without the remnant.
 In the original work \cite{Bowick:1988xh, Bowick:1989xg}, it was suggested that the wormhole solution, the  instanton solution for gravity coupled to $B_2$   \cite{Giddings:1987cg}, can be taken into account in addition.
More concretely, the tunneling process in the wormhole solution is interpreted as the nucleation of the baby universe.
 Since the PQ charge can flow into this baby universe, the PQ charge in our universe is not conserved \cite{Coleman:1988cy, Giddings:1988cx, Lee:1988ge, Rey:1989mg}, which provides the mechanism to reduce the PQ charge of the black hole.
  
  On the other hand, \cite{Hebecker:2017uix} and \cite{Montero:2017yja} proposed another process, in which the string `lassoing' the black hole and interacting with $B_2$ is introduced.
  Similarly to the wormhhole solution, the effective potential generated by the string is given by $V(q)=e^{-4\pi r^2 {\cal T}_2}\cos q$, where $4\pi r^2$ is the worldsheet area with $r$ given by the  radius of the black hole and ${\cal T}_2$ is the string tension.
  From this, the PQ charge $q$ can be treated as the dynamical variable and stabilized at $q=0$ by the potential.
 As pointed out in \cite{Hebecker:2017uix}, for this mechanism to work, the condition $r \lesssim 1/\sqrt{\cal T}_2$ needs to be satisfied such that the instanton factor $e^{-4\pi r^2 {\cal T}_2}$ in $V(q)$ is not significantly suppressed.
Moreover, for the complete evaporation of black hole to be consistent with the energy conservation, the energy carried by the non-vanishing $H_3$ (the leftover field strength after the complete evaporation) must be smaller than the black hole mass $\sim r M_{\rm Pl}^2$.
Thus, we expect the inequality
 \dis{E=\frac{1}{f^2}\int \frac12 H_3\wedge *_4 H_3 \sim \frac{q^2}{f^2 r^3} \lesssim  r M_{\rm Pl}^2 \label{eq:Econs}}
 is satisfied.
 Combining this with the condition  $r \lesssim 1/\sqrt{\cal T}_2$, we obtain \cite{Hebecker:2017uix}
 \dis{{\cal T}_2 \lesssim \frac{f}{q}M_{\rm Pl}. \label{eq:T21}}
 Applying the axion WGC-like bound \eqref{eq:aWGC}, this gives the upper bound on the tension of the string,
 \dis{{\cal T}_2 \lesssim \frac{1}{q} \frac{g^2}{8\pi^2}M_{\rm Pl}^2.\label{eq:T22}}
 We note that $q$ is the PQ charge whereas $g$ is the gauge coupling.
 
  To see how this bound on ${\cal T}_2$ is satisfied in   string models, we first consider the model independent axion in the heterotic string theory.
 We  infer from \eqref{eq:hetf} and \eqref{eq:hetg} (or \eqref{eq:Hetf}) that the fundamental string tension  ${\cal T}_{{\rm F}1}=\frac{1}{2\pi \alpha'}=\frac{2\pi}{\ell_s^2}$  can be rewritten as
 \dis{{\cal T}_{{\rm F}1} = 2\pi \frac{8\pi^2}{g^2} f^2 = \frac{1}{\sqrt2}(2\pi f)M_{\rm Pl}=\pi \frac{g^2}{8\pi^2} M_{\rm Pl}^2,}
 which  saturates the bound on ${\cal T}_2$ given by \eqref{eq:T21} as well as \eqref{eq:T22} for $q\sim {\cal O}(1)$. 
 Meanwhile, for the axion $\theta^\alpha$ from $C_4$, the PQ charge  can be carries by the string obtained from wrapping the D$3$-brane on a 2-cycle  $\Sigma_2$ satisfying $\int_{\Sigma_2}b^\alpha = n^\alpha$, where $b^\alpha$ is the basis of the harmonic $(1,1)$-form and the winding number $n^\alpha$ is an integer.
  Since the K\"ahler form $\omega$ satisfies $\int_{\Sigma_2} \omega=n^\alpha t_\alpha$ and the tension of the D$3$-brane is given by $\frac{2\pi}{g_s\ell_s^4}$, the tension of the string is 
  \dis{{\cal T}_2 = \frac{2\pi}{g_s\ell_s^4}{\rm Vol}(\Sigma_2)=\frac{2\pi}{g_s\ell_s^2}n^\alpha t_\alpha.}
 In order to rewrite ${\cal T}_2$ in terms of  $f$ (or $M_{\rm Pl}$) as above, we use the relations $t_\alpha=\partial_{\tau^\alpha}\partial_{\tau^\beta}{\cal K}\tau^\beta{\cal V}$, $g_\alpha^2=\frac{2\pi g_s}{\tau^\alpha}$, and $M_{\rm Pl}^2=\frac{4\pi{\cal V}}{g_s^2 \ell_s^2}$,   to obtain
 \dis{  f^2_{\alpha\beta}\frac{8\pi^2}{g_\beta^2}= \frac{g_s^2}{8\pi^2}M_{\rm Pl}^2 \frac{\partial^2 {\cal K}}{\partial \tau^\alpha \partial{\tau^\beta}}\times \tau^\beta\frac{8\pi^2}{2\pi g_s} =\frac{2 t_\alpha}{g_s\ell_s^2}.}
 Comparing this with ${\cal T}_2$, we find the relation similar to that in the model independent axion case :
 \dis{{\cal T}_2=\pi n^\alpha f_{\alpha\beta}^2 \frac{8\pi^2}{g_\beta^2}.}
 Using  \eqref{eq:C4f}, we also finds that
 \dis{\frac{4}{\pi}\frac{{\cal T}_2}{M_{\rm Pl}^2}&= g_{\alpha}^2 n^\alpha \frac{\partial^2 {\cal K}}{\partial {\tau^\alpha}\partial\tau^\beta } \tau^\alpha\tau^\beta =\sum_\alpha g_\alpha^2 n^\alpha \frac{t_\alpha\tau^\alpha}{\cal V}
 \\
 &\leq{\rm max}(g_\alpha^2 n^\alpha)\sum_\alpha \frac{t_\alpha\tau^\alpha}{\cal V}= 3\times{\rm max}(g_\alpha^2 n^\alpha). } 
 In our concrete example in which the volume is given by \eqref{eq:LVS}, we obtain from   \eqref{eq:fcheese} and \eqref{eq:fcheese2} that
 \dis{\frac{4}{\pi}\frac{{\cal T}_2}{M_{\rm Pl}^2} =&g_5^2 n^{(5)}\tau_5^2\Big[-\frac{\sqrt2}{12}\frac{1}{{\cal V}\tau_5^{1/2}}+\frac{1}{36}\frac{\tau_5}{{\cal V}^2}\Big]
 +g_4^2 n^{(4)}\tau_4^2 \Big[\frac{\sqrt2}{12}\frac{1}{{\cal V}\tau_4^{1/2}}+\frac{1}{36}\frac{\tau_4}{{\cal V}^2}\Big]
 \\
 &+(g_5^2n^{(5)}+g_4^2n^{(4)})\tau_5\tau_4\Big[-\frac{1}{36}\frac{\tau_5^{1/2}\tau_4^{1/2}}{{\cal V}^2}\Big].}
This shows that unless $g_5$ and $n^{(5)}$ are   much smaller than $g_4$ and $n^{(4)}$, the first term is dominant to give ${\cal T}_2\sim g_5^2 M_{\rm Pl}^2$ as $\tau_5\sim {\cal V}^{2/3}$ and $\tau_4\sim \log{\cal V}$.
 Moreover, for $g_5=g_4$ and $n^{(5)}=n^{(4)}$, ${\cal T}_2=\frac{3\pi}{4}g_5^2 n^{(5)}M_{\rm Pl}^2 $ is satisfied, as expected.
In any case, the bound on  ${\cal T}_2$ given by \eqref{eq:T22} is saturated.
On the other hand, when $n^{(5)}=0$, i.e., the string does not couple to the axion $\theta^{(5)}$, the tension is suppressed as ${\cal T}_2 \sim \frac{(\log{\cal V})^{3/2}}{\cal V}$.
 
 
 We point out here that whereas the strings we considered, the fundamental string or the D$3$-brane wrapping the 2-cycle, play the crucial role in reducing the PQ charge of the black hole, they are not present in the EFT.
 That is, for the strings to be excited, we need the energy well above the cutoff scale.
 Moreover, our discussions on the (axion) species scale suggest  that the appropriate cutoff scale of the EFT might be  $\Lambda_{\rm sp}$ or $\Lambda_{{\rm sp}, a}$, rather than $M_{\rm Pl}$ or $\frac{8\pi^2}{g^2}f$.
 This is also supported by the fact that the (axion) species scale for   the string tower is given by the string mass scale, i.e., $\Lambda_{\rm sp} \simeq \Lambda_{{\rm sp}, a} \simeq \ell_s^{-1}$, above which   we can naturally find the string excitations.
 
 To see the implication of it, we note from the energy conservation condition \eqref{eq:Econs} that 
  the minimal radius of the axion black hole is given by
 \dis{  r_{\rm min}\equiv\frac{1}{ \sqrt{f M_{\rm Pl}}}.}   
 Intriguingly, in the wormhole solution, the same length scale $\sim \frac{1}{ \sqrt{f M_{\rm Pl}}}$ is interpreted as the size of the wormhole throat \cite{Giddings:1987cg}.
 \footnote{One caveat here is that, at the core of the wormhole, dynamics of the `saxion' (or radial) mode is no longer ignorable (see, e.g., section 5 of \cite{Hebecker:2016dsw}), which may provide  the model dependent correction to the throat size. } 
 Thus, we expect that even though the length scale $r_{\rm min}$ is longer than $M_{\rm Pl}^{-1}$ as super-Planckian   physics is not taken into account to obtain it, gravity   is already strong enough that the PQ symmetry breaking effects are no longer negligible.
 Indeed, we have found that $r_{\rm min}$ is   shorter than $1/\sqrt{{\cal T}_2}$ so  the interaction between black hole and the string is quite active.
 In addition, at this scale, the PQ charge begins to flow into the baby universe through the wormhole throat.
 From these, we may require that $r_{\rm min}$ is similar to or shorter than $\Lambda_{\rm sp}^{-1}=\sqrt{N_{\rm sp}}/M_{\rm Pl}$, which reads
\dis{f \gtrsim \frac{\Lambda_{\rm sp}}{\sqrt{N_{\rm sp}}}=\frac{M_{\rm Pl}}{N_{\rm sp}}=M_{\rm Pl} \Big(\frac{m_{t}}{M_{\rm Pl}}\Big)^{\frac{2 p}{p+2}}.\label{eq:BoundM}}
For the string tower ($m_t=\ell_s^{-1}$ and $p\to\infty$), the RHS is given by $(M_{\rm Pl}\ell_s^2)^{-1}$.  
Combining with the axion WGC-like bound  $f \lesssim \frac{g^2}{8\pi^2}M_{\rm Pl}$,  the inequality above sets the lower bound on the gauge coupling,
\dis{\frac{g^2}{8\pi^2} \gtrsim \frac{1}{N_{\rm sp}}=\Big(\frac{m_{t}}{M_{\rm Pl}}\Big)^{\frac{2 p}{p+2}}.\label{eq:BoundM2}}

On the other hand, the axion WGC-like bound was  obtained by imposing $\Lambda_{{\rm sp},a} \lesssim \Lambda_{\rm sp}$, thus the condition $r_{\rm min} \lesssim \Lambda_{\rm sp}^{-1}$ automatically leads to $r_{\rm min}\lesssim \Lambda_{{\rm sp},a}^{-1}$.
This means that even though the energy scale at which  the PQ symmetry is broken   is low compared to $\frac{8\pi^2}{g^2}f$, the effects from the  $\theta F\wedge F$ term cannot be completely ignored  due to the coupling between the axion and a tower of states.
 We can rewrite the condition $r_{\rm min}\lesssim \Lambda_{{\rm sp},a}^{-1}$ as
\dis{\Big(\frac{8\pi^2}{g^2}\Big)^2 f \lesssim N_{{\rm sp}, a}M_{\rm Pl}.}
This shows that similarly to \eqref{eq:BoundM2}, the value of $\frac{g^2}{8\pi^2}$ is larger than $N_{{\rm sp}, a}^{-1}$ times the ratio $\frac{8\pi^2}{g^2}\frac{f}{M_{\rm Pl}}$ which is smaller than $1$. 
 When the axion WGC-like bound  is saturated, i.e., $\frac{8\pi^2}{g^2}\simeq \frac{M_{\rm Pl}}{f}$, the bound becomes
 \dis{\frac{g^2}{8\pi^2} \gtrsim \frac{1}{N_{{\rm sp}, a}}, \label{eq:BoundM3}}
 which is more stringent than \eqref{eq:BoundM2}  since $N_{{\rm sp}, a}$ cannot exceed $N_{{\rm sp}}$.
In particular, for the string tower ($m_s=\ell_s^{-1}$ and $p \to \infty$),  
\dis{N_{\rm sp}\simeq N_{{\rm sp}, a}\simeq \Big(\frac{M_{\rm Pl}}{m_s}\Big)^2 =\frac{4\pi {\cal V}}{g_s^2},}
where \eqref{eq:hetg} is used for the last equality, is satisfied, and $g^2$ is given by $g_s$.
Then the condition \eqref{eq:BoundM2} (with $m_s=\ell_s^{-1}$ and $p \to \infty$) as well as \eqref{eq:BoundM3} reads $g_s \gtrsim \frac{g_s^2}{4\pi{\cal V}}$, which is trivial   in the parametrically controllable regime satisfying $g_s<1$ and ${\cal V}>1$.
On the other hand, it may be notable that the gauge coupling is given by $\frac{g^2}{8\pi^2}=\frac{g_s^2}{2\pi {\cal V}}$ (see \eqref{eq:hetg}), thus in the presence of the string tower $\frac{g^2}{8\pi^2} \simeq \frac{1}{N_{\rm sp}}$ is satisfied for the model independent axion in the heterotic string theory.
For the axion from $C_4$, putting $\frac{g_\alpha^2}{8\pi^2}=\frac{g_s}{4\pi\tau^\alpha}$ (see \eqref{eq:g2C4}) into \eqref{eq:BoundM2} gives $\frac{\tau^\alpha}{\cal V} g_s \lesssim 1$.
This is also trivial    due to the string perturbativity $g_s <1$ and the fact that the volume of the 4-cycle is smaller than the overall volume ($\tau^\alpha < {\cal V}$). 

 Now we further investigate the implication of \eqref{eq:BoundM2} and \eqref{eq:BoundM3}   from the thermodynamic point of view.
 Since the number of states made up of   $N_{\rm sp}$ species  is estimated as $N_{\rm sp}!$, we can define the `species entropy' by   \cite{Cribiori:2023ffn, Basile:2024dqq, Herraez:2024kux}
 \dis{\log N_{\rm sp}! \sim N_{\rm sp}  \log N_{\rm sp} \sim N_{\rm sp}.}
 If the species entropy can be interpreted as the entropy of the low energy EFT, it must be smaller than the entropy of the universe : 
 \dis{N_{\rm sp} \lesssim \frac{M_{\rm Pl}^2}{H^2},}
 where $H$ is the Hubble parameter, from which the cosmological constant is given by $3M_{\rm Pl}^2H^2$.
 Using the same argument, we also obtain  $N_{{\rm sp}, a}\lesssim \frac{M_{\rm Pl}^2}{H^2}$ as well.
 Combining these with \eqref{eq:BoundM2} and \eqref{eq:BoundM3}, we arrive at
 \dis{\frac{g^2}{8\pi^2}\gtrsim \frac{H^2}{M_{\rm Pl}^2}.\label{eq:ComBo}}
 Meanwhile, for the U(1) gauge interaction, there is another conjectured bound relating   $g$ and $H$ given by 
 \dis{\frac{g^2}{8\pi^2}\lesssim \frac{m^4}{H^2 M_{\rm Pl}^2},\label{eq:FLb}}
 where $m$ is the mass of the U(1) charged particle.
 This is called the Festina-Lente bound \cite{Montero:2019ekk}, which can be justified by the cosmic censorship hypothesis \cite{Penrose:1969pc} : in the process of the U(1) charged black hole evaporation in de Sitter space, the singularity must not be naked. 
Then  the bounds \eqref{eq:ComBo} and \eqref{eq:FLb} indicate   $m\gtrsim H$, i.e., the mass of the U(1) charged particle must not exceed the Hubble parameter.
\footnote{If the argument above can be extended to the non-Abelian gauge interaction such as the weak interaction, the bound $m\gtrsim H$ is remarkable since there have been attempts to explain the neutrino mass $m\sim H$ in terms of the swampland conjectures \cite{Ibanez:2017kvh, Hamada:2017yji, Gonzalo:2021zsp, Casas:2024clw}.}

 \section{Conclusions}
\label{sec:conclusion}

In this article, we  investigate the connection between the axion WGC-like bound and the distance conjecture  by considering the axion  species scale, the energy scale at which the perturbativity breaks down by the interaction between the  axion and towers of states.
It becomes clear that   the axion WGC-like bound $\frac{8\pi^2}{g^2} f\lesssim M_{\rm Pl}$  can be interpreted as the condition that the axion species scale cannot exceed the species scale. 
The concrete string theory models show the existence of the axion decay constant saturating the axion WGC-like bound, as already well known.
At the energy scale above the (axion) species scale, we expect that  the PQ symmetry breaking effects from quantum gravity  are no longer negligible.
Comparing the characteristic length scale $\frac{1}{\sqrt{ f M_{\rm Pl}}}$ of the PQ symmetry breaking with the (axion) species scale leads to the lower bound on the gauge coupling or the string coupling given by the inverse of the number of species in the EFT.

 We may extend our discussions on the axion species scale and the axion WGC-like bound to the generic non-renormalizable interactions.
 That is, for any non-renormalizable interaction, we can always find the cutoff scale above which the perturbativity breaks down.
 Since this cutoff scale typically depends on the number of   species in the EFT, we can call it the species scale associated with the non-renormalizable interaction.
 The distance conjecture suggests that 
  the species scale can be much lower than the characteristic scale of the non-renormalizable interaction when  the particles in the EFT interact with a tower of states through the corresponding non-renormalizable interaction.
 Whereas we may postulate the various UV completions of the non-renormalizable interaction, so far as quantum gravity is the most microscopic theory, the species scale associated with the non-renormalizable interaction is expected to be lower than the (gravitational) species scale.
 This sets the upper bound on the characteristic scale of the non-renormalizable interaction depending on the Planck scale and the coupling, which has the same form as the axion WGC bound.
 In particular, when the non-renormalizable interaction originates from quantum gravity, we expect that the bound can be saturated and presumably, the concrete string model can be found.


%


\appendix


\renewcommand{\theequation}{\Alph{section}.\arabic{equation}}



\begin{thebibliography}{99}

\small

\bibitem{Kim:1986ax}
J.~E.~Kim,
Phys. Rept. \textbf{150} (1987), 1-177

\bibitem{Kim:2008hd}
J.~E.~Kim and G.~Carosi,
Rev. Mod. Phys. \textbf{82} (2010), 557-602
[erratum: Rev. Mod. Phys. \textbf{91} (2019) no.4, 049902]
[arXiv:0807.3125 [hep-ph]].

\bibitem{Peccei:1977ur}
R.~D.~Peccei and H.~R.~Quinn,
Phys. Rev. D \textbf{16} (1977), 1791-1797

\bibitem{Peccei:1977hh}
R.~D.~Peccei and H.~R.~Quinn,
Phys. Rev. Lett. \textbf{38} (1977), 1440-1443

\bibitem{Weinberg:1977ma}
S.~Weinberg,
Phys. Rev. Lett. \textbf{40} (1978), 223-226

\bibitem{Wilczek:1977pj}
F.~Wilczek,
Phys. Rev. Lett. \textbf{40} (1978), 279-282


\bibitem{Vafa:1984xg}
C.~Vafa and E.~Witten,
Phys. Rev. Lett. \textbf{53} (1984), 535

\bibitem{Kim:1979if}
J.~E.~Kim,
Phys. Rev. Lett. \textbf{43} (1979), 103

\bibitem{Shifman:1979if}
M.~A.~Shifman, A.~I.~Vainshtein and V.~I.~Zakharov,
Nucl. Phys. B \textbf{166} (1980), 493-506

\bibitem{Zhitnitsky:1980tq}
A.~R.~Zhitnitsky,
Sov. J. Nucl. Phys. \textbf{31} (1980), 260

\bibitem{Dine:1981rt}
M.~Dine, W.~Fischler and M.~Srednicki,
Phys. Lett. B \textbf{104} (1981), 199-202

\bibitem{Witten:1984dg}
E.~Witten,
Phys. Lett. B \textbf{149} (1984), 351-356


\bibitem{Svrcek:2006yi}
P.~Svrcek and E.~Witten,
JHEP \textbf{06} (2006), 051
[arXiv:hep-th/0605206 [hep-th]].

\bibitem{Burgess:1998px}
C.~P.~Burgess, L.~E.~Ibanez and F.~Quevedo,
Phys. Lett. B \textbf{447} (1999), 257-265
[arXiv:hep-ph/9810535 [hep-ph]].

\bibitem{Choi:2003wr}
K.~w.~Choi,
Phys. Rev. Lett. \textbf{92} (2004), 101602
[arXiv:hep-ph/0308024 [hep-ph]].

\bibitem{Conlon:2006tq}
J.~P.~Conlon,
JHEP \textbf{05} (2006), 078
[arXiv:hep-th/0602233 [hep-th]].

\bibitem{Choi:2014uaa}
K.~Choi, K.~S.~Jeong and M.~S.~Seo,
JHEP \textbf{07} (2014), 092
[arXiv:1404.3880 [hep-th]].

\bibitem{Gendler:2024gdo}
N.~Gendler and C.~Vafa,
[arXiv:2404.15414 [hep-th]].

\bibitem{Arkani-Hamed:2003xts}
N.~Arkani-Hamed, H.~C.~Cheng, P.~Creminelli and L.~Randall,
Phys. Rev. Lett. \textbf{90} (2003), 221302
[arXiv:hep-th/0301218 [hep-th]].

\bibitem{Svrcek:2006hf}
P.~Svrcek,
[arXiv:hep-th/0607086 [hep-th]].

\bibitem{Arvanitaki:2009fg}
A.~Arvanitaki, S.~Dimopoulos, S.~Dubovsky, N.~Kaloper and J.~March-Russell,
Phys. Rev. D \textbf{81} (2010), 123530
[arXiv:0905.4720 [hep-th]].


\bibitem{Choi:2020rgn}
K.~Choi, S.~H.~Im and C.~S. Shin,
Ann. Rev. Nucl. Part. Sci. \textbf{71} (2021), 225-252
[arXiv:2012.05029 [hep-ph]].

\bibitem{Vafa:2005ui}
C.~Vafa,
[arXiv:hep-th/0509212 [hep-th]].

\bibitem{Brennan:2017rbf}
T.~D.~Brennan, F.~Carta and C.~Vafa,
PoS \textbf{TASI2017} (2017), 015
[arXiv:1711.00864 [hep-th]].

\bibitem{Palti:2019pca}
E.~Palti,
Fortsch. Phys. \textbf{67} (2019) no.6, 1900037
[arXiv:1903.06239 [hep-th]].

\bibitem{vanBeest:2021lhn}
M.~van Beest, J.~Calder\'on-Infante, D.~Mirfendereski and I.~Valenzuela,
Phys. Rept. \textbf{989} (2022), 1-50
[arXiv:2102.01111 [hep-th]].

\bibitem{Grana:2021zvf}
M.~Gra\~na and A.~Herr\'aez,
Universe \textbf{7} (2021) no.8, 273
[arXiv:2107.00087 [hep-th]].

\bibitem{Agmon:2022thq}
N.~B.~Agmon, A.~Bedroya, M.~J.~Kang and C.~Vafa,
[arXiv:2212.06187 [hep-th]].

\bibitem{VanRiet:2023pnx}
T.~Van Riet and G.~Zoccarato,
Phys. Rept. \textbf{1049} (2024), 1-51
[arXiv:2305.01722 [hep-th]].

\bibitem{Arkani-Hamed:2006emk}
N.~Arkani-Hamed, L.~Motl, A.~Nicolis and C.~Vafa,
JHEP \textbf{06} (2007), 060
[arXiv:hep-th/0601001 [hep-th]].

\bibitem{Harlow:2022ich}
D.~Harlow, B.~Heidenreich, M.~Reece and T.~Rudelius,
Rev. Mod. Phys. \textbf{95} (2023) no.3, 3
[arXiv:2201.08380 [hep-th]].

\bibitem{Banks:2006mm}
T.~Banks, M.~Johnson and A.~Shomer,
JHEP \textbf{09} (2006), 049
[arXiv:hep-th/0606277 [hep-th]].

\bibitem{Reece:2023czb}
M.~Reece,
PoS \textbf{TASI2022} (2024), 008
[arXiv:2304.08512 [hep-ph]].

\bibitem{Rudelius:2014wla}
T.~Rudelius,
JCAP \textbf{04} (2015), 049
[arXiv:1409.5793 [hep-th]].

\bibitem{Rudelius:2015xta}
T.~Rudelius,
JCAP \textbf{09} (2015), 020
[arXiv:1503.00795 [hep-th]].

\bibitem{Montero:2015ofa}
M.~Montero, A.~M.~Uranga and I.~Valenzuela,
JHEP \textbf{08} (2015), 032
[arXiv:1503.03886 [hep-th]].

\bibitem{Brown:2015iha}
J.~Brown, W.~Cottrell, G.~Shiu and P.~Soler,
JHEP \textbf{10} (2015), 023
[arXiv:1503.04783 [hep-th]].

\bibitem{Hebecker:2015rya}
A.~Hebecker, P.~Mangat, F.~Rompineve and L.~T.~Witkowski,
Phys. Lett. B \textbf{748} (2015), 455-462
[arXiv:1503.07912 [hep-th]].

\bibitem{Brown:2015lia}
J.~Brown, W.~Cottrell, G.~Shiu and P.~Soler,
JHEP \textbf{04} (2016), 017
[arXiv:1504.00659 [hep-th]].

\bibitem{Heidenreich:2015wga}
B.~Heidenreich, M.~Reece and T.~Rudelius,
JHEP \textbf{12} (2015), 108
[arXiv:1506.03447 [hep-th]].




\bibitem{Ooguri:2006in}
H.~Ooguri and C.~Vafa,
Nucl. Phys. B \textbf{766} (2007), 21-33
[arXiv:hep-th/0605264 [hep-th]].

\bibitem{Lee:2019xtm}
S.~J.~Lee, W.~Lerche and T.~Weigand,
JHEP \textbf{02} (2022), 096
[arXiv:1904.06344 [hep-th]].

\bibitem{Lee:2019wij}
S.~J.~Lee, W.~Lerche and T.~Weigand,
JHEP \textbf{02} (2022), 190
[arXiv:1910.01135 [hep-th]].

\bibitem{Veneziano:2001ah}
G.~Veneziano,
JHEP \textbf{06} (2002), 051
[arXiv:hep-th/0110129 [hep-th]].

\bibitem{Dvali:2007hz}
G.~Dvali,
Fortsch. Phys. \textbf{58} (2010), 528-536
[arXiv:0706.2050 [hep-th]].

\bibitem{Dvali:2007wp}
G.~Dvali and M.~Redi,
Phys. Rev. D \textbf{77} (2008), 045027
[arXiv:0710.4344 [hep-th]].

\bibitem{Dvali:2009ks}
G.~Dvali and D.~Lust,
Fortsch. Phys. \textbf{58} (2010), 505-527
[arXiv:0912.3167 [hep-th]].

\bibitem{Dvali:2010vm}
G.~Dvali and C.~Gomez,
[arXiv:1004.3744 [hep-th]].

\bibitem{Dvali:2012uq}
G.~Dvali, C.~Gomez and D.~Lust,
Fortsch. Phys. \textbf{61} (2013), 768-778
[arXiv:1206.2365 [hep-th]].

\bibitem{Reece:2024wrn}
M.~Reece,
[arXiv:2406.08543 [hep-ph]].

\bibitem{Heidenreich:2017sim}
B.~Heidenreich, M.~Reece and T.~Rudelius,
Eur. Phys. J. C \textbf{78} (2018) no.4, 337
[arXiv:1712.01868 [hep-th]].

\bibitem{Bowick:1988xh}
M.~J.~Bowick, S.~B.~Giddings, J.~A.~Harvey, G.~T.~Horowitz and A.~Strominger,
Phys. Rev. Lett. \textbf{61} (1988), 2823

\bibitem{Bowick:1989xg}
M.~J.~Bowick,
Gen. Rel. Grav. \textbf{22} (1990), 137-144

\bibitem{Hebecker:2017uix}
A.~Hebecker and P.~Soler,
JHEP \textbf{09} (2017), 036
[arXiv:1702.06130 [hep-th]].

\bibitem{Montero:2017yja}
M.~Montero, A.~M.~Uranga and I.~Valenzuela,
JHEP \textbf{07} (2017), 123
[arXiv:1702.06147 [hep-th]].

\bibitem{Giddings:1987cg}
S.~B.~Giddings and A.~Strominger,
Nucl. Phys. B \textbf{306} (1988), 890-907

\bibitem{Hebecker:2016dsw}
A.~Hebecker, P.~Mangat, S.~Theisen and L.~T.~Witkowski,
JHEP \textbf{02} (2017), 097
[arXiv:1607.06814 [hep-th]].

\bibitem{Hebecker:2018ofv}
A.~Hebecker, T.~Mikhail and P.~Soler,
Front. Astron. Space Sci. \textbf{5} (2018), 35 
[arXiv:1807.00824 [hep-th]].

\bibitem{Montero:2019ekk}
M.~Montero, T.~Van Riet and G.~Venken,
JHEP \textbf{01} (2020), 039
[arXiv:1910.01648 [hep-th]].


\bibitem{Montero:2021otb}
M.~Montero, C.~Vafa, T.~Van Riet and G.~Venken,
JHEP \textbf{10} (2021), 009
[arXiv:2106.07650 [hep-th]].

\bibitem{Lee:2021cor}
S.~M.~Lee, D.~Y.~Cheong, S.~C.~Hyun, S.~C.~Park and M.~S.~Seo,
JHEP \textbf{02} (2022), 100
[arXiv:2111.04010 [hep-ph]].

\bibitem{Blumenhagen:2013fgp}
R.~Blumenhagen, D.~L\"ust and S.~Theisen,
``Basic concepts of string theory,''
Springer, 2013.


\bibitem{Castellano:2022bvr}
A.~Castellano, A.~Herr\'aez and L.~E.~Ib\'a\~nez,
JHEP \textbf{06} (2023), 047
[arXiv:2212.03908 [hep-th]].
 
\bibitem{Calmet:2017omb}
X.~Calmet, R.~Casadio, A.~Y.~Kamenshchik and O.~V.~Teryaev,
Phys. Lett. B \textbf{774} (2017), 332-337
[arXiv:1708.01485 [hep-th]].

\bibitem{Castellano:2021mmx}
A.~Castellano, A.~Herr\'aez and L.~E.~Ib\'a\~nez,
JHEP \textbf{08} (2022), 217
[arXiv:2112.10796 [hep-th]].



\bibitem{Kani:1989im}
I.~Kani and C.~Vafa,
Commun. Math. Phys. \textbf{130} (1990), 529-580

\bibitem{Seo:2023xsb}
M.~S.~Seo,
JHEP \textbf{09} (2023), 031
[arXiv:2305.18673 [hep-th]].

\bibitem{Fan:2021ntg}
J.~Fan, K.~Fraser, M.~Reece and J.~Stout,
Phys. Rev. Lett. \textbf{127} (2021) no.13, 131602
[arXiv:2105.09950 [hep-ph]].

\bibitem{Denef:2004dm}
F.~Denef, M.~R.~Douglas and B.~Florea,
JHEP \textbf{06} (2004), 034
[arXiv:hep-th/0404257 [hep-th]].

\bibitem{Candelas:1994hw}
P.~Candelas, A.~Font, S.~H.~Katz and D.~R.~Morrison,
Nucl. Phys. B \textbf{429} (1994), 626-674
[arXiv:hep-th/9403187 [hep-th]].

\bibitem{Balasubramanian:2005zx}
V.~Balasubramanian, P.~Berglund, J.~P.~Conlon and F.~Quevedo,
JHEP \textbf{03} (2005), 007
[arXiv:hep-th/0502058 [hep-th]].





\bibitem{Susskind:1995da}
L.~Susskind,
[arXiv:hep-th/9501106 [hep-th]].




\bibitem{Coleman:1988cy}
S.~R.~Coleman,
Nucl. Phys. B \textbf{307} (1988), 867-882

\bibitem{Giddings:1988cx}
S.~B.~Giddings and A.~Strominger,
Nucl. Phys. B \textbf{307} (1988), 854-866

\bibitem{Lee:1988ge}
K.~M.~Lee,
Phys. Rev. Lett. \textbf{61} (1988), 263-266

\bibitem{Rey:1989mg}
S.~J.~Rey,
Phys. Rev. D \textbf{39} (1989), 3185

\bibitem{Cribiori:2023ffn}
N.~Cribiori, D.~Lust and C.~Montella,
JHEP \textbf{10} (2023), 059
[arXiv:2305.10489 [hep-th]].

\bibitem{Basile:2024dqq}
I.~Basile, N.~Cribiori, D.~Lust and C.~Montella,
JHEP \textbf{06} (2024), 127
[arXiv:2401.06851 [hep-th]].

\bibitem{Herraez:2024kux}
A.~Herr\'aez, D.~L\"ust, J.~Masias and M.~Scalisi,
[arXiv:2406.17851 [hep-th]].



\bibitem{Penrose:1969pc}
R.~Penrose,
Riv. Nuovo Cim. \textbf{1} (1969), 252-276

\bibitem{Ibanez:2017kvh}
L.~E.~Ibanez, V.~Martin-Lozano and I.~Valenzuela,
JHEP \textbf{11} (2017), 066
[arXiv:1706.05392 [hep-th]].

\bibitem{Hamada:2017yji}
Y.~Hamada and G.~Shiu,
JHEP \textbf{11} (2017), 043
[arXiv:1707.06326 [hep-th]].

\bibitem{Gonzalo:2021zsp}
E.~Gonzalo, L.~E.~Ib\'a\~nez and I.~Valenzuela,
JHEP \textbf{02} (2022), 088
[arXiv:2109.10961 [hep-th]].

\bibitem{Casas:2024clw}
G.~F.~Casas, L.~E.~Ib\'a\~nez and F.~Marchesano,
[arXiv:2406.14609 [hep-th]].

\end{thebibliography}
\end{document}